\documentclass[review]{elsarticle}

\usepackage{hyperref}
\usepackage{amsmath}
\usepackage{array}
\usepackage{graphicx}
\usepackage{epstopdf}
\usepackage{amssymb}
\usepackage{float}
\usepackage{enumerate}
\usepackage{booktabs}
\usepackage{enumitem}
\usepackage{balance}
\usepackage{xr}
\usepackage{subcaption}
\usepackage{booktabs}
\usepackage{adjustbox}
\usepackage{multirow}
\usepackage{multicol}

\journal{Journal of Pattern Recognition}

\begin{document}

\begin{frontmatter}

\title{Longitudinal Prediction of Postnatal Brain Magnetic Resonance Images via a Metamorphic Generative Adversarial Network}


\author[mainaddress,secondaryaddress]{Yunzhi~Huang}
\author[secondaryaddress]{Sahar~Ahmad}
\author[thirdaddress]{Luyi~Han}
\author[fourthaddress]{Shuai~Wang}
\author[secondaryaddress]{Zhengwang~Wu}
\author[secondaryaddress]{Weili~Lin}
\author[secondaryaddress]{Gang~Li}
\author[secondaryaddress]{Li~Wang}
\author[secondaryaddress]{Pew-Thian Yap\corref{correspondingauthor}}
\cortext[correspondingauthor]{Corresponding author}
\ead{ptyap@med.unc.edu}

\address[mainaddress]{School of Automation, Nanjing University of Information Science and Technology, Nanjing 210044, China}
\address[secondaryaddress]{Department of Radiology and Biomedical Research Imaging Center (BRIC), University of North Carolina, Chapel Hill, USA}
\address[thirdaddress]{Department of Radiology and Nuclear Medicine, Radboud University Medical Center, Geert Grooteplein 10, 6525 GA, Nijmegen, The Netherlands}
\address[fourthaddress]{Department of Computer Science, Shandong University (Weihai), China}

\begin{abstract}
Missing scans are inevitable in longitudinal studies due to either subject dropouts or failed scans. 
In this paper, we propose a deep learning framework to predict missing scans from acquired scans, catering to longitudinal infant studies. Prediction of infant brain MRI is challenging owing to the rapid contrast and structural changes particularly during the first year of life. We introduce a trustworthy metamorphic generative adversarial network (MGAN) for translating infant brain MRI from one time-point to another. 
MGAN has three key features: (i) Image translation leveraging spatial and frequency information for detail-preserving mapping; (ii) Quality-guided learning strategy that focuses attention on challenging regions. (iii) Multi-scale hybrid loss function that improves translation of tissue contrast and structural details. 
Experimental results indicate that MGAN outperforms existing GANs by accurately predicting both contrast and anatomical details.
\end{abstract}


\end{frontmatter}


\section{Introduction}
\label{introduction}
Brain MRI is commonly used to investigate normative and aberrant brain evolution through infancy~\cite{dubois2008mapping}. 
To precisely chart brain growth trajectories, \emph{temporally dense} longitudinal datasets are often required but are difficult to acquire. Moreover, infant studies often involve incomplete longitudinal datasets, given the unique challenges associated with infant MRI acquisition. The missing data at different time points can be due to subject dropouts or failed scans owing to excessive motion, insufficient coverage, or imaging artifacts~\cite{howell2019TheUNC}.

\begin{figure}[tbp]
	\centerline{\includegraphics[width=0.8\textwidth]{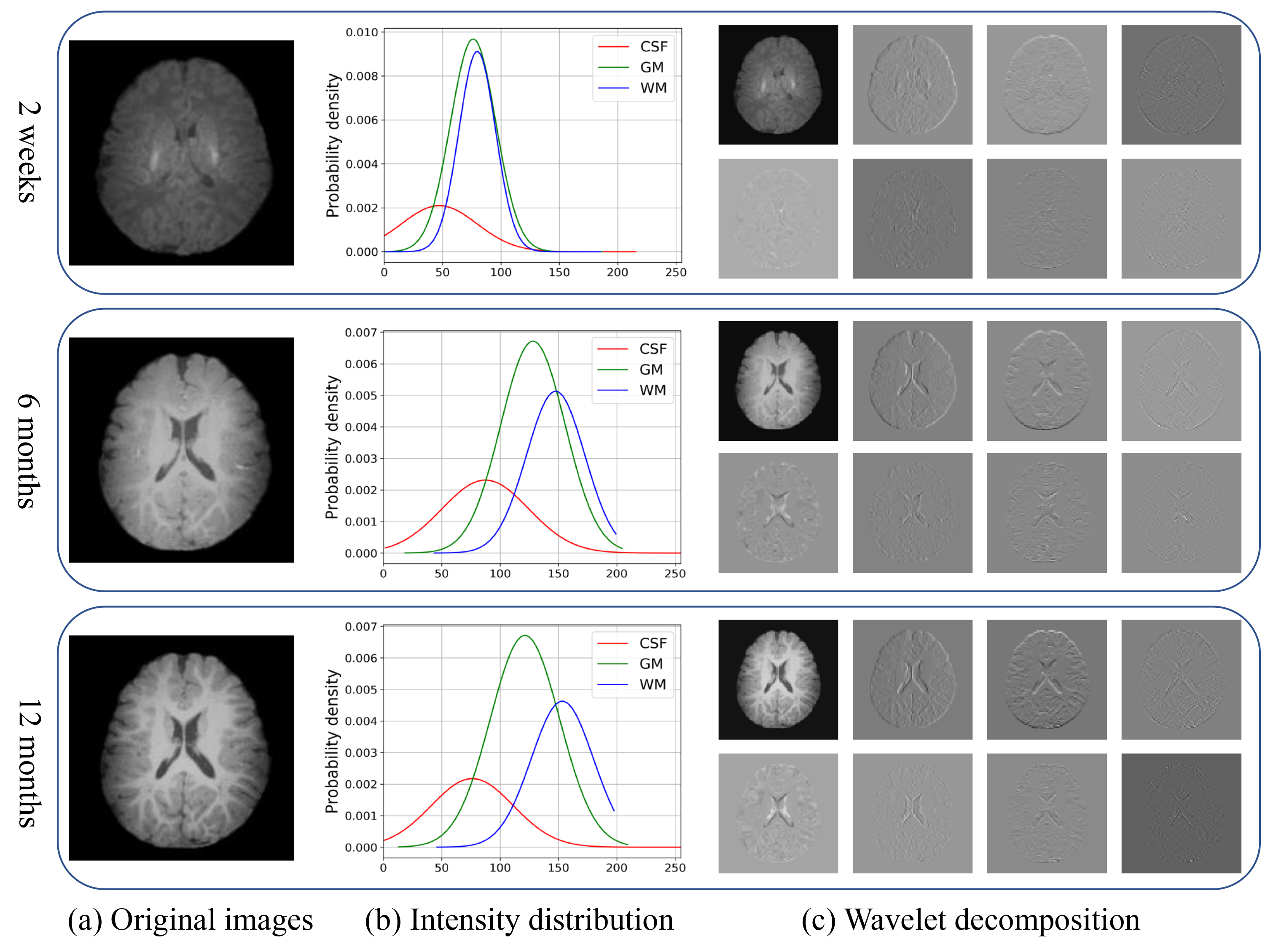}}
	\caption{Appearance and structural changes at two time points during the first year of life. Wavelet decomposition for capturing structural details.}
	\label{fig:background}
\end{figure}

Longitudinal prediction of infant brain scans is challenging as brain MRI contrasts change rapidly through the first year of life. The brain volume doubles to about $65\%$ of the adult brain by the end of the first year~\cite{gilmore2018imaging}. The gray matter (GM) follows a faster growth trajectory ($108\% - 149\%$ increase) compared to white matter (WM; $11\%$ increase)~\cite{Matsuzawa2001Age}. 
The rapid brain evolution is characterized by both structural and contrast variations~\cite{knickmeyer2008structural,paus2001maturation}. As shown in Figure~\ref{fig:background}, the WM appears to be darker than the GM during the neonatal phase as the brain is going through myelination, and by sixth month, WM and GM are almost indistinguishable due to the poor tissue contrast. 

\subsection{Related Work}
The longitudinal prediction of infant brain MRI can be formulated as an image-to-image translation task --- mapping images from a source time point to a target time point~\cite{Isola2016ImagetoImageTW}. Several studies in the field of computer vision~\cite{Isola2016ImagetoImageTW,Zhu2017UnpairedIT,Huang2018MultimodalUI,Liu2017UnsupervisedIT} have shown that generative adversarial networks (GANs)~\cite{Goodfellow2014GenerativeAN} yield superior performance in translating images from a domain to another.
In the field of medical image analysis,  
\cite{nie2017medical} introduced an auto-context GAN to progressively refine MRI-to-CT synthesis.  
In their follow-up study, \cite{nie2020adversarial} incorporated difficulty-aware attention mechanism to improve predictions in challenging regions. Similarly, \cite{Lee2018DavinciGANUS} introduced self-attention to encourage the transformation of a foreground object while retaining the background.
Medical image-to-image translation network (MedGAN)~\cite{Armanious2019MedGANMI} uses a pre-trained classification network as feature extractor to match textures and structural details of synthetic and target CT images. 
All the aforementioned methods for cross-modality synthesis focus on appearance changes and neglect morphological changes. The longitudinal prediction of infant MR brain images, however, requires dealing with fast-paced structural and appearance changes.

To promote structural consistency in cross-modality synthesis, several recent approaches incorporate segmentation similarity as a learning constraint~\cite{Zhang2018TranslatingAS,Chartsias2017AdversarialIS}.
However, tissue segmentation of infant brain MRI is challenging due to the overlap of GM and WM intensity distributions (Fig.~\ref{fig:background}). 
Several approaches attempted to ensure structural consistency without relying on tissue maps. \cite{nie2018medical} employed gradient differences in a loss function to improve the prediction of boundaries. \cite{hiasa2018cross} incorporated gradient correlation differences in a structure-consistency loss to improve edge alignment in MRI-to-CT synthesis. Although  successful, the gradient-based constraint introduces noise and fail to capture sufficient boundary information in images with low contrast.
\cite{Yang2018UnpairedBM} incorporated a patch-based self-similarity loss by comparing each patch with all its neighbors in a pre-defined non-local region to ensure structural consistency. However, the search for corresponding non-local regions is computationally expensive.

\subsection{Contributions}
In this paper, we employ CycleGAN~\cite{Zhu2017UnpairedIT}, a cycle consistent generation framework, to simultaneously learn structural and appearance changes between two time points. Major contributions of our work are summarized below:
\begin{enumerate}[label=(\roman*)]
	\item We propose a trustworthy adversarial learning metamorphosis framework that accounts for both the appearance and structural changes in infant brain MRI. 
	\item We use a spatial-frequency transfer block equipped with wavelet decomposition to transform features from multiple frequency bands to learn the structural changes.
	\item We employ a quality guidance strategy to incorporate a quality-driven loss function to improve predictions in challenging regions.
	\item We devise a multi-scale hybrid loss function to improve the matching of both the textural details and the anatomical edges between the predicted image and the desired target image. The discriminator network is evoked at multiple resolutions via deep-supervision, thus allowing accurate prediction of anatomical structures through adversarial learning. 
\end{enumerate}

The rest of the paper is organized as follows: Section~\ref{sec:Methods} details the proposed method. Section~\ref{sec:results} describes the dataset used for evaluation and presents the experimental results. Section~\ref{sec:discussion} provides additional discussion and concludes the paper.

\section{Methods}
\label{sec:Methods}
In this work, we implement a framework for prediction of metamorphic changes using a GAN. 
Details of our method are described next.

\subsection{Network Architecture}
\begin{figure*}[!t]
	\centering
	\includegraphics[width=\textwidth]{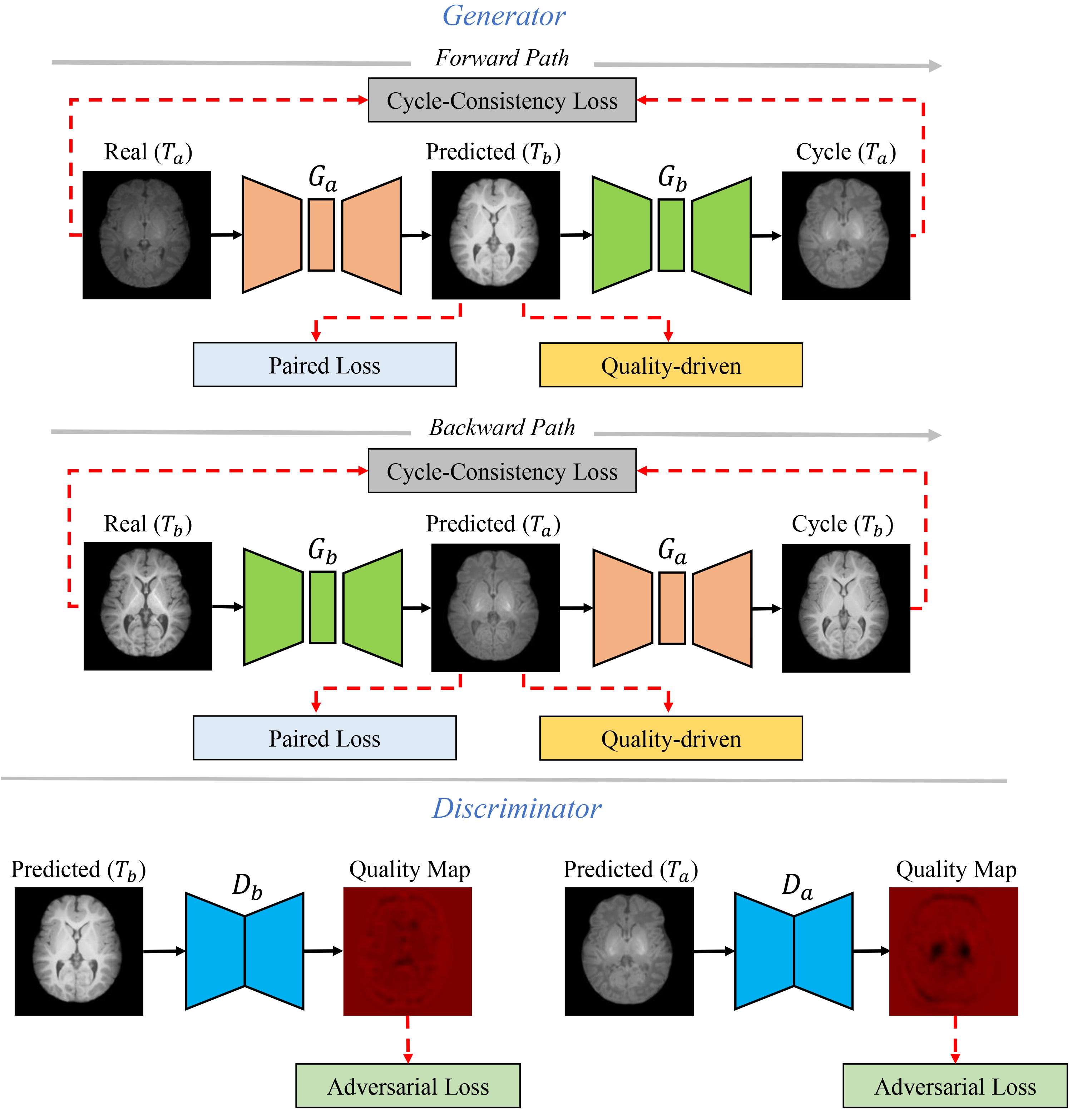}
	\caption{Overview of the metamorphic GAN.}
	\label{fig:overview}
\end{figure*}

We propose a metamorphic GAN (MGAN) to predict the infant brain MR image scanned at time point $t_{b}$ from a time point $t_{a}$. Without loss of generality, we assume that $t_{b}>t_{a}$.
Our network architecture, shown in Fig.~\ref{fig:overview}, is cycle-consistent and learns a reversible translation between the two time-points. It consists of (i) a forward path for earlier-to-later time-point image prediction and (ii) a backward path for later-to-earlier time-point image prediction. The two generators $G_{a}$ and $G_{b}$ and their corresponding discriminators $D_{a}$ and $D_{b}$ follow an encoder-decoder architecture.
Both the generators incorporate a spatial-frequency transfer (SFT) block to transform the appearance and structural features via multiple branches detailed in Fig.~\ref{fig:generator}.
The two discriminators estimate voxel-level uncertainty maps, enabling the corresponding generators to focus on challenging regions. We will describe the components of our network in the subsequent sections.

\subsubsection{Metamorphic Generator}
\begin{figure*}[!htbp]
	\centering
	\includegraphics[width=0.95\textwidth]{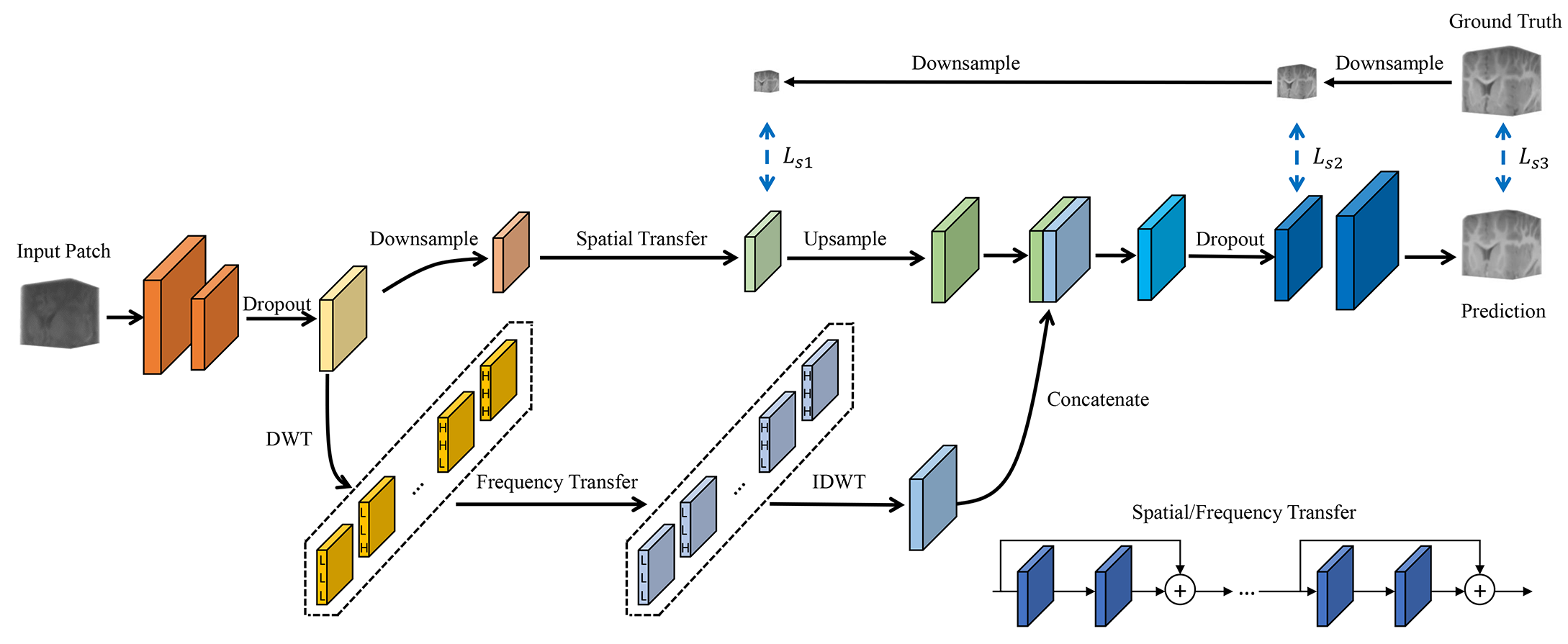}
	\caption{Network architecture of the metamorphic generator.}
	\label{fig:generator}
\end{figure*}

The metamorphic generator (Fig.~\ref{fig:generator}) takes a 3D patch of size $64\times64\times64$ as input and predicts a 3D patch. The generator consists of an encoder, SFT block, and a decoder.

\paragraph{Encoder}
The encoding path consists of two convolution blocks, each with a $3\times3\times3$ convolution layer, followed by $3$D instance normalization (IN)~\cite{Ulyanov2016InstanceNT} and a rectified linear unit (ReLU)~\cite{Glorot2011DeepSR}.  
For downsampling, we use convolution with a stride of $2$ instead of pooling to avoid potential information loss. We keep a $1$-stride convolution in the first stage of the encoder to retain details, and use a $2$-stride convolution in the second stage. The resulting numbers of feature maps in the two-stage encoder are $64$ and $32$. 

\paragraph{Spatial-frequency transfer block}
Longitudinal prediction requires translating both contrast and structure between two time points. We propose to embed a spatial-frequency transfer block in between enocder-decoder to extract the spatial and frequency domain information of feature maps. The SFT block is divided into two branches: (i) frequency transform branch, and (ii) spatial transform branch. The frequency transform branch is equipped with discrete wavelet transform (DWT) that takes into account the low frequency tissue contrast and high frequency structural details. The DWT layer decomposes the feature map into low frequency approximation and high frequency details along three dimensions, resulting in eight subvolumes: $LLL, LLH, LHL, LHH, HLL, HLH, HHL$ and $HHH$. This decomposition allows more effective transfer of spatial-frequency details.

\begin{figure}[!htbp]
	\centering
	\includegraphics[width=0.8\textwidth]{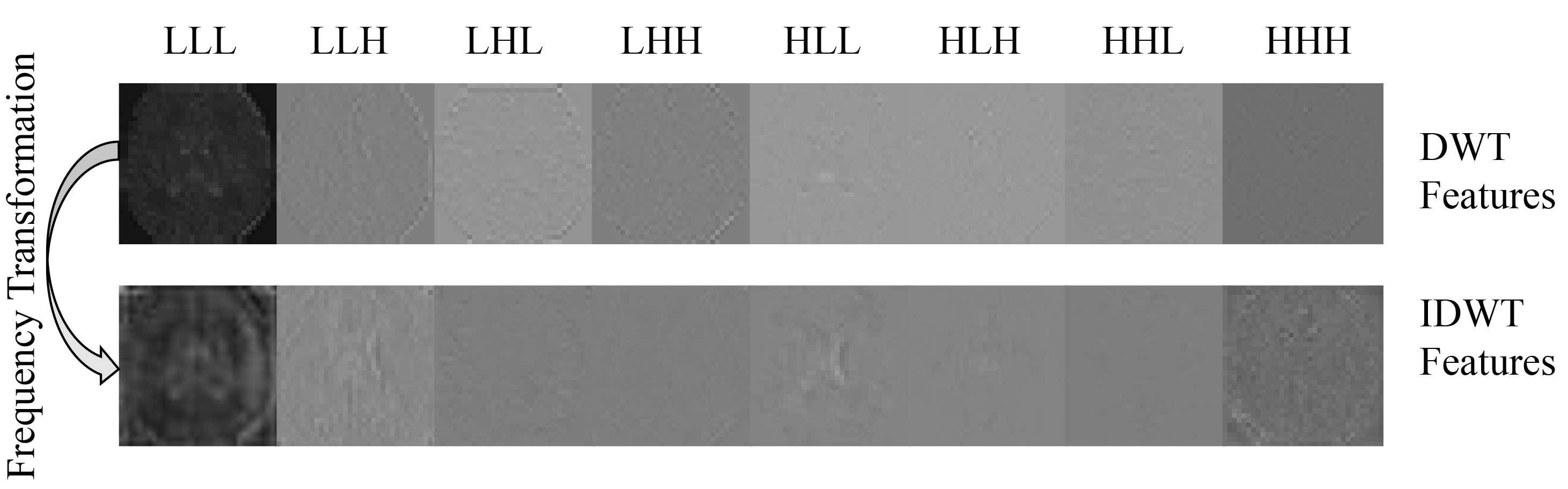}
	\caption{Feature maps obtained from the DWT and IDWT layers.}
	\label{fig:wavelet}
\end{figure}

Given the $i$-th channel feature map $f^{i}$ of size $(s_{x}\times s_{y}\times s_{z})$, the decomposed feature map $f^{i}_{j}$ at frequency band $j$ is obtained by convolving $f^{i}$ with wavelet filter $w_j$:
\begin{equation}
\label{eqa:dwt}
f_{j}^{i}=f^{i}\circledast w_{j}.
\end{equation}
The wavelet filters for each frequency band are calculated by DWT decomposition and are preset in the convolution layer. Correspondingly, the feature maps are reconstructed in the decode path via inverse discrete wavelet transform (IDWT) layer. We show the representative feature maps from the DWT and IDWT layers in Fig.~\ref{fig:wavelet}. The DWT layer is akin to pooling layer as the DWT decomposition halves the size of the input feature maps. The IDWT layer corresponds to the deconvolution operation with the fixed weights obtained via wavelet filters. There is also an intermediate \emph{transfer} operation between the DWT and IDWT layer. This transfer operation is realized through $9$ residual blocks~\cite{He2016DeepRL}; the input of each block is processed by two $3\times3\times3$ convolution layers with $64$ channels followed by IN and ReLU for activation. A shortcut connection is added between the input and the output of every residual convolution block. Residual transfer learning simplifies feature generation and transfer from a source domain to a target domain. 	

The second branch in the SFT block --- spatial transform branch --- is integrated to compensate for the information truncated by the wavelets. It is implemented using a convolution layer with kernel size $3\times3\times3$ and stride of $2$ to downsample the feature maps in spatial domain. These downsampled feature maps undergo transfer operation and are later upsampled by strided deconvolution layer with kernel size $3\times3\times3$.

Each branch in the SFT block is trained independently without weight sharing. The feature maps from both the frequency and spatial transform branches are concatenated using a $3\times3\times3$ convolution layer and stride of $1$, followed by IN and ReLU operation; capturing both the contrast and structural information for translating from the source domain to the target domain. 

\paragraph{Decoder} 
Deep supervision~\cite{Xie2015HolisticallyNestedED} is leveraged in the decoding path to strengthen the gradient flow and encourage learning useful representations at multiple scales. The feature maps are upsampled by a $2$-stride deconvolution layer and are then convolved with a $3\times3\times3$ kernel to get the predicted output.

\subsubsection{Uncertainty Quantization} 
The uncertainty associated with the prediction stems from two aspects: \textit{epistemic} uncertainty (model uncertainty) and \textit{aleatoric} uncertainty (data uncertainty)~\cite{der2009aleatory,abdar2020review}. 
As shown in Fig.~\ref{fig:generator}, two Monte-Carlo (MC) dropout layers are incorporated in our generator to estimate the epistemic uncertainty.
MC dropout regularizes the network weights as \textit{Bernoulli} distributions for variational Bayesian inference~\cite{kendall2015bayesian,gal2016dropout}.
Note, MC dropout is only enabled during inference. A set of predictions $\{\hat{y}_{1},\hat{y}_{2},\dots,\hat{y}_{N}\}$ are sampled from the distribution $p(\hat{y}|I,\mathbf{w}_{n})$ via $N$ stochastic inferences using the metamorphic generator. The epistemic uncertainty is estimated as the variance over the predictions:
\begin{equation}
\label{eq:modelUN}
\mathcal{U}_{e}=\sqrt{\dfrac{\sum_{n=1}^{N}(\hat{y}_{n}-\overline{y})^{2}}{N}},
\end{equation}	
where $\hat{y}$ denotes to the prediction by feeding the generator $G$ an input image $I$, $\mathbf{w}_{n}$ represents the generator weights after the $n$-th dropout, $N$ refers to the number of prediction instances, and $\overline{y}$ is the mean of the predictions.

The aleatoric uncertainty is typically measured with the test-time augmentation technique~\cite{ayhan2018test,wang2019aleatoric}. 
During inference, we perturb the input data with spatial transformations (flip and rotation) and random noise. Similar to the estimation of the epistemic uncertainty, we sample a set of predictions $\{\hat{y}_{1},\hat{y}_{2},\dots,\hat{y}_{N}\}$ from the distribution $p(\hat{y}|I,S))$ and estimate the aleatoric uncertainty as the variance over the predictions:
\begin{equation}
\label{eq:dataUN}
\mathcal{U}_{a} = \sqrt{\dfrac{\sum_{n=1}^{N}(S^{-1}(\hat{y}(S(x+rn))-\overline{y})^{2}}{N}},
\end{equation}
where $S$ represents the spatial transformation, $S^{-1}$ corresponds to the inverse transformation, and $rn$ corresponds to random noise. 


\subsubsection{Multi-scale discriminator}
\begin{figure}[!htbp]
	\centering
	\includegraphics[width=\textwidth]{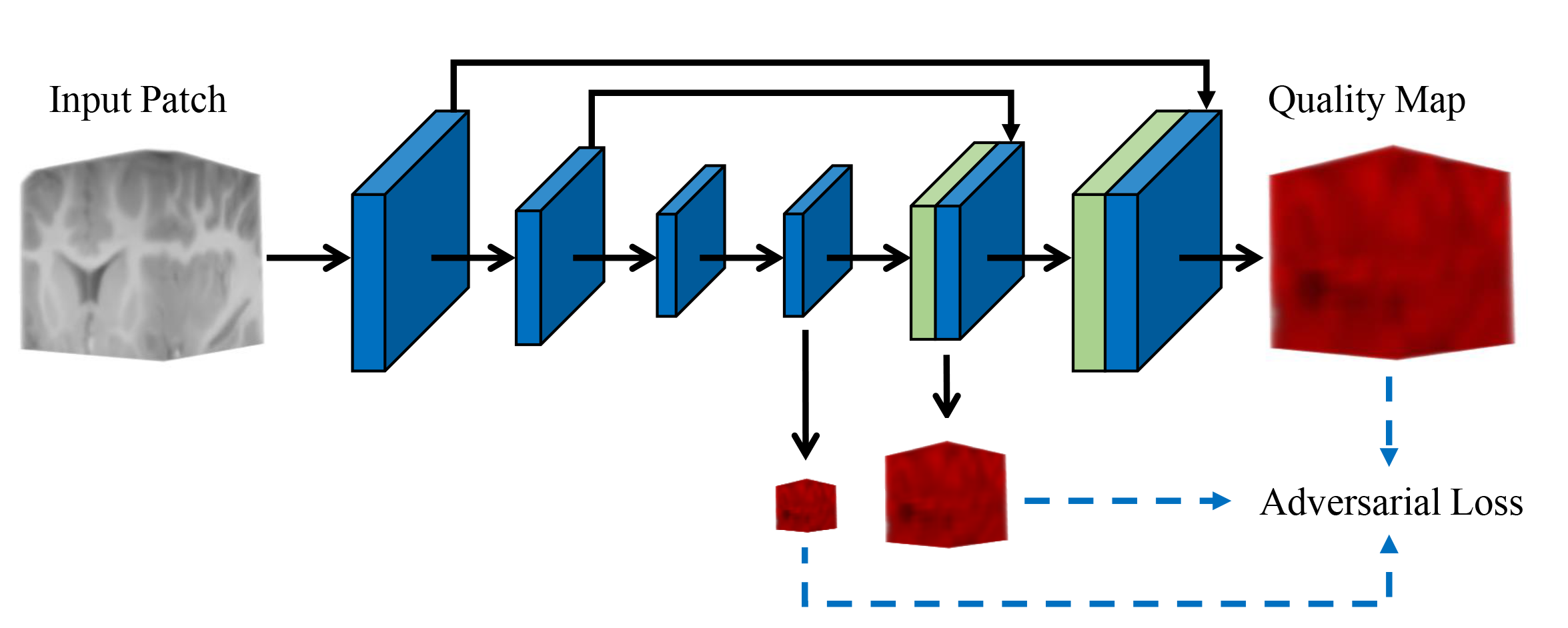}
	\caption{Network architecture of the multi-scale discriminator.}
	\label{fig:discriminator}
\end{figure}

The discriminator in MGAN has a U-shaped architecture, as shown in Fig.~\ref{fig:discriminator}, to locally distinguish the predicted images from real images. It takes as input a $64\times64\times64$ image patch and outputs the quality probability map for the given $3$D patch. The continuous probability map quantifies the quality of the predicted image patch. Inferior quality, associated with lower probability values, is commonly associated with complex structures, e.g., the cortical ribbon. Superior quality, associated with higher probability values, corresponds to flat regions with simple structures.
In the encoding path of the discriminator, the input is downsampled three times; in the decoding path, the feature maps are upsampled three times. For downsampling/upsampling, we use a $4\times4\times4$ convolution/deconvolution layer, followed by IN and ReLU activation. The numbers of feature channels are $64$, $128$, and $256$ in the three stages of the discriminator. Deep supervision strategy~\cite{Xie2015HolisticallyNestedED} is incorporated in the decoding path to strengthen gradient back propagation.

\subsection{Loss Functions}
We incorporated supervised learning with multi-scale information via deep supervision strategy~\cite{Xie2015HolisticallyNestedED}. The loss function $\mathcal{L}_{\mathrm{MGAN}}$ is defined as:
\begin{equation}
\label{eq:totalloss}
\mathcal{L}_{\mathrm{MGAN}} = \mathcal{L}_{s_{1}}+\mathcal{L}_{s_{2}}+\mathcal{L}_{s_{3}},
\end{equation}
\noindent where $s_{1}$, $s_{2}$, and $s_{3}$ refer to the three scales employed~\cite{wang2018high,karnewar2020msg}. For each scale, the objective function is composed of three loss functions to effectively learn the prediction task. The loss functions are described next.

\subsubsection{Adversarial Loss}
We propose to use the standard adversarial loss function, which aims to match the distribution of the predicted images with that of the real images. It is given by
\begin{equation}
\label{eq:gan}
\begin{aligned}
\mathcal{L}(G_{a},G_{b},D_{a},D_{b})
& = \mathbb{E}_{I_{t_{a}}}[\log(D_{a}(I_{t_{a}})]\\ 
& + \mathbb{E}_{I_{t_{b}}}[\log(1-D_{a}(G_{b}(I_{t_{b}})))]\\
& + \mathbb{E}_{I_{t_{b}}}[\log(D_{b}(I_{t_{b}})]\\ 
& + \mathbb{E}_{I_{t_{a}}}[\log(1-D_{b}(G_{a}(I_{t_{a}})))],
\end{aligned}
\end{equation}
where $I_{t_{a}}$ and $I_{t_{b}}$ refer to the images at time-point $t_{a}$ and $t_{b}$, respectively, $G_{a}$ and $G_{b}$ are the mapping functions, and $D_{a}$ and $D_{b}$ are the discriminators.

\subsubsection{Paired Loss}
The generators $G_{a}$ and $G_{b}$ seek to minimize the difference between real and predicted images. We propose to enhance the performance of the generators by defining a paired loss function that constraints the difference at voxel-, feature-, and frequency-level. Our paired loss function $\mathcal{L}_{\mathrm{gen}}^{(\cdot)}$ consists of three loss terms: (i) quality-driven loss, (ii) texture loss, and (iii) frequency loss.
\begin{equation}
\label{eq:gen}
\begin{aligned}
\mathcal{L}_{\mathrm{gen}}^{G_{a}}&=\mathcal{L}_{\mathrm{Q}}^{G_{a}}+\mathcal{L}_{\mathrm{T}}^{G_{a}} +\mathcal{L}_{\mathrm{F}}^{G_{a}}, \\
\mathcal{L}_{\mathrm{gen}}^{G_{b}}&=\mathcal{L}_{\mathrm{Q}}^{G_{b}}+\mathcal{L}_{\mathrm{T}}^{G_{b}} +\mathcal{L}_{\mathrm{F}}^{G_{b}}.
\end{aligned}
\end{equation}

\paragraph{Quality-driven loss}
The low tissue contrast and the dramatic brain growth hinder translation of regions such as the convoluted cerebral cortex. Here, we present a quality-guided learning strategy to strengthen the transformation of the unfathomable regions. The discriminator outputs a quality map that defines the voxel-wise probabilities for each predicted image. The heterogeneous distribution of the probabilities in the quality map motivates us to treat voxels differently. Voxels with lower probabilities correspond to poor prediction and require more attention compared to those with higher probabilities. This enhances the image translation power of the generator at complex regions in the infant brain MRI. The quality-driven loss $\mathcal{L}_{\mathrm{Q}}^{(\cdot)}$ is defined as:
\begin{equation}
\label{eq:quality}
\begin{aligned}
\mathcal{L}_{\mathrm{Q}}^{G_{a}}(G_{a};\theta^{G_{a}})=\mathbb{E}_{I_{t_{a}},I_{t_{b}},Q^{D_{b}}}[\|I_{t_{b}}-G_{a}(I_{t_{a}})\|_{1}\odot(1-Q^{D_{b}})^\beta],\\
\mathcal{L}_{\mathrm{Q}}^{G_{b}}(G_{b};\theta^{G_{b}})=\mathbb{E}_{I_{t_{a}},I_{t_{b}},Q^{D_{a}}}[\|I_{t_{a}}-G_{b}(I_{t_{b}})\|_{1}\odot(1-Q^{D_{a}})^\beta],
\end{aligned}
\end{equation}
where $Q^{(\cdot)}$ is the quality map, $\theta^{(\cdot)}$ denotes the parameters of the network, $\odot$ defines the element-wise multiplication and $\beta$ represents the parameter that enables to focus on difficult-to-predict regions. If $\beta$ is set to zero, then $\mathcal{L}_{\mathrm{Q}}^{(\cdot)}$ will be equivalent to $\mathcal{L}_{1}$ norm; losing the ability to define adaptive weights based on quality map. In this study, we empirically set its value to $1.5$.

\paragraph{Texture loss}
This loss ensures that the predicted image has a texture similar to the target image, and it is defined as the mean square error (MSE) between the Gram matrix of the target and the predicted image~\cite{gatys2015texture,li2017diversified}:
\begin{equation}
\label{eq:gram}
\begin{aligned}
\mathcal{L}_{\mathrm{T}}^{G_{a}}&=\Arrowvert M(I_{t_{b}})-M(G_{a}(I_{t_{a}})) \|_2,\\
\mathcal{L}_{\mathrm{T}}^{G_{b}}&=\Arrowvert M(I_{t_{a}})-M(G_{b}(I_{t_{b}})) \|_2.
\end{aligned}
\end{equation}
The gram matrix $M(\cdot)$ is the inner product of the generated images.

\paragraph{Frequency loss}	
The frequency loss $\mathcal{L}_{\mathrm{F}}^{(\cdot)}$ is incorporated via wavelet decomposition of the generators' outputs, which steers the effective prediction of the structural details. $\mathcal{L}_{\mathrm{F}}^{(\cdot)}$ is defined as:
\begin{equation}
\label{eq:wave}
\begin{aligned}
\mathcal{L}_{\mathrm{F}}^{G_{a}}&=\sum_{k \in K}\Arrowvert \mathrm{DWT}(I_{t_{b}})^{k}-\mathrm{DWT}(G_{a}(I_{t_{a}}))^{k} \lVert_1, \\
\mathcal{L}_{\mathrm{F}}^{G_{b}}&=\sum_{k \in K}\Arrowvert \mathrm{DWT}(I_{t_{a}})^{k}-\mathrm{DWT}(G_{b}(I_{t_{b}}))^{k} \lVert_1,
\end{aligned}
\end{equation}
where $K = \allowbreak\{LLL,LLH,LHL,HLL,LHH,HLH,HHL,\allowbreak HHH\}$; $LLL$ corresponds to the approximation coefficients which encode the image contrast, and the remaining terms correspond to the detail coefficients, encoding the high frequency structural details. The wavelet coefficients are decomposed using bior1.3~\cite{Cohen2006BiorthogonalBO}, which is compactly supported by a biorthogonal spline wavelet~\cite{szewczyk2012reliable}. 

\subsubsection{Cycle Consistency Loss}
The cycle consistency loss function $\mathcal{L}_{\mathrm{cyc}}$ ensures that the image prediction cycle brings the predicted image back to the original image, i.e., $G_{b}(G_{a}(I_{t_{a}})) \approx I_{t_{a}}$ and it is given by:
\begin{equation}
\label{eq:cyc}
\begin{aligned}
\mathcal{L}_{\mathrm{cyc}}(G_{a},G_{b})
&= \mathbb{E}_{I_{t_{a}}}[\Arrowvert I_{t_{a}} - G_{b}(G_{a}(I_{t_{a}}))\lVert_{1}],\\ 
&+ \mathbb{E}_{I_{t_{b}}}[\Arrowvert I_{{t_{b}}} - G_{a}(G_{b}(I_{{t_{b}}}))\lVert_{1}].
\end{aligned}
\end{equation}
This loss function constraints both the forward and backward image prediction cycles, causing $G_{a}$ and $G_{b}$ to be consistent with each other.	

\section{Experimental Results}
\subsection{Data Acquisition and Preprocessing}
\label{sec:data}
The dataset consists of longitudinal T1-weighted (T1w) and T2-weighted (T2w) MR images of healthy infant subjects enrolled in the Multi-visit Advanced Pediatric Brain Imaging (MAP) study. Informed written consent was obtained from the parents of all the participants and all study protocols were approved by the University of North Carolina at Chapel Hill Institutional Review Board. Each subject was scanned every three months in the first postnatal year. The imaging parameters for T1w MRI data were: TR $=1900$ ms, TE $=4.38$ ms, flip angle $=7^{\circ}$. All the images had $144$ sagittal slices and $1$\,mm isotropic voxel resolution. The imaging parameters for T2w MR images were TR = 7380 ms, TE = 119 ms, flip angle = $150^{\circ}$, $64$ sagittal slices, and $1.25\times1.25\times1.95\,\text{mm}^3$ voxel size.

The dataset was preprocessed using our infant-dedicated preprocessing pipeline~\cite{dai2013ibeat,li2013mapping}.
Then, all the postnatal images of each subject were linearly aligned to their corresponding 12-months-old images and resampled to the size of $256\times256\times256$ with $1\times1\times1\,\mathrm{mm}^{3}$ voxel resolution. 
We randomly split the MRI data from $30$ healthy infants into 20 and 10 for training and testing, respectively. Five-fold cross-validation was performed to tune the hyper-parameters.

\subsection{Implementation Details}
The proposed metamorphic GAN was implemented using TensorFlow library~\cite{Abadi2016TensorFlow} on a single Nvidia TitanX (Pascal) GPU. Adam optimizer~\cite{Kingma2015AdamAM} was adopted with an initial learning rate of $1\times10^{-4}$ and batch size of $1$. Training, validation, and testing were performed separately for T1w and T2w images.

During training, we uniformly sampled $3$D patches from each image encompassing the brain region with a dense stride of 10, providing sufficient samples for training. The generator was first trained with $5$ epochs before the adversarial training. The adversarial training was stopped at $50$ epochs. 

During inference, the $N=20$ inferences were performed for the estimation of epistemic and aleatoric uncertainty. The keep rate of the dropout layers was set to $0.8$. Test-time data augmentation was carried out using a combination of random flip, rotation along each of the three axes, and random noise, which were modeled respectively with discrete Bernoulli distribution $\mathcal{B}(0.5)$, uniform distribution $\mathcal{U}(0,2\pi)$, and normal distribution $\mathcal{N}(0,0.05)$.

\label{sec:results}
\subsection{Evaluation Criteria}
We employed two commonly used metrics to evaluate the quality of the predicted images: (i) peak signal-to-noise ratio (PSNR), and (ii) structural similarity (SSIM)~\cite{Wang2004ImageQA}. 
Higher PSNR and SSIM correspond to accurate image prediction.

\subsection{Comparison with Existing Techniques}
We compared MGAN with three widely used GANs: CycleGAN~\cite{Zhang2018TranslatingAS}, Pix2Pix~\cite{Isola2016ImagetoImageTW}, and WGAN~\cite{Arjovsky2017WassersteinGA}. All the compared models were used to predict the 12-month-old brain MRI from the 2-week-old brain MRI. The prediction task is challenging due to the extent of changes between the two time points (Fig.~\ref{fig:background}). For fair comparison, we re-trained the GANs for optimal parameters. 

\begin{figure*}[!htbp]
	\begin{adjustbox}{center}
		\begin{subfigure}{1.1\textwidth}
			\includegraphics[width=\linewidth]{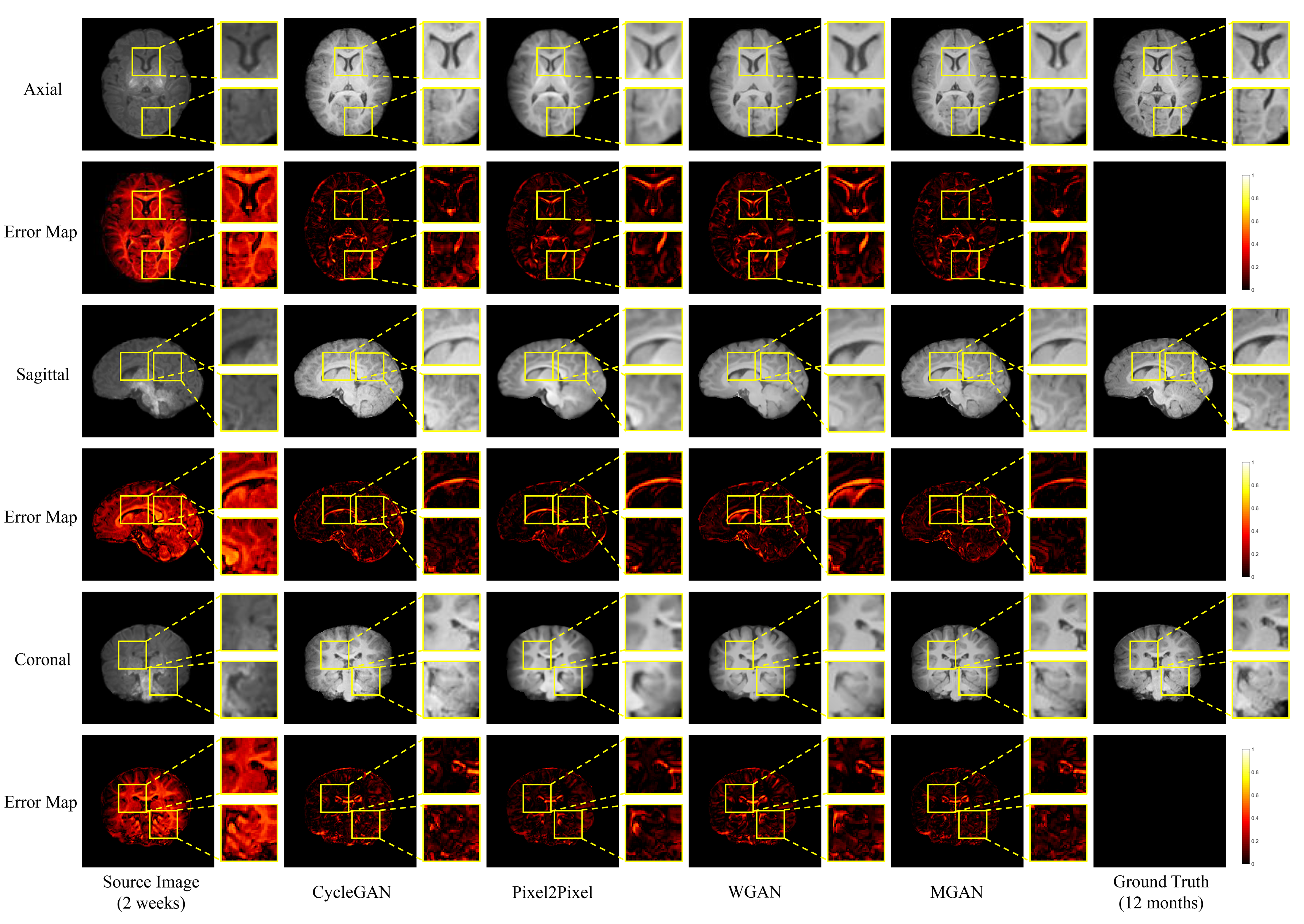}
			\caption{T1w image predictions}
			\label{fig:alternativesT1}
		\end{subfigure}
	\end{adjustbox}
	
	\vspace{\baselineskip}
	
	\begin{adjustbox}{center}
		\begin{subfigure}{1.1\textwidth}
			\centering
			\includegraphics[width=\linewidth]{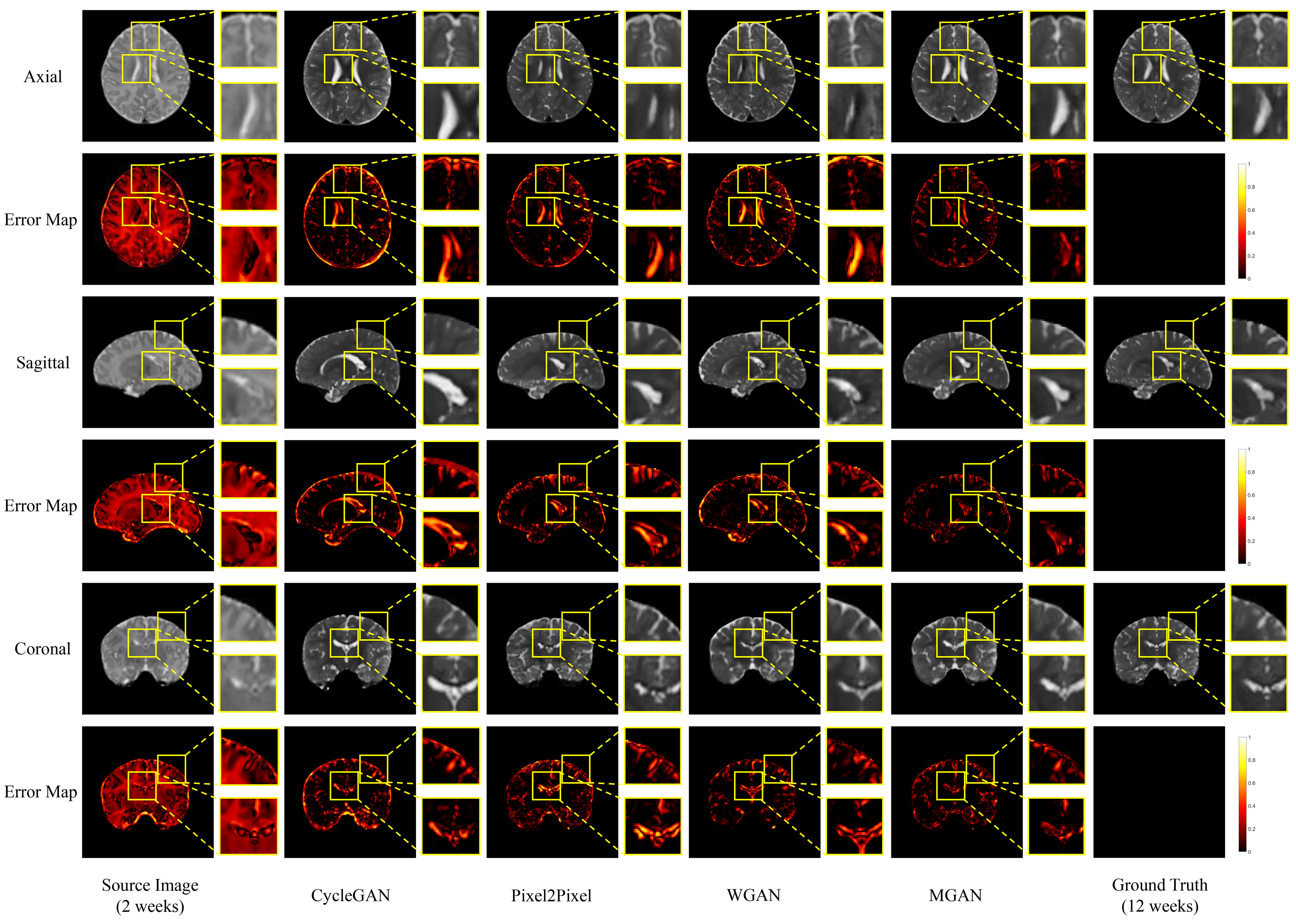}
			\caption{T2w image predictions}
			\label{fig:alternativesT2}
		\end{subfigure}
	\end{adjustbox}
	\caption{Longitudinal image prediction with various GANs.}\label{fig:alternatives}
\end{figure*}

The image prediction results shown for the compared models in Fig.~\ref{fig:alternatives} indicate that MGAN yields T1w and T2w image predictions that are closer to the ground truth with richer details than the other models. The error maps indicate that MGAN achieves the lowest error among all methods, especially around the ventricles and cerebral cortex. 
Summary statistics for PSNR and SSIM are reported in Table~\ref{tab:alternatives}. MGAN achieves significant improvement ($p<0.05$, paired $t$-test) for PSNR and SSIM over other methods. 

\begin{table}[!htpb]
	\renewcommand{\arraystretch}{1.3}
	\centering
	\caption{Summary statistics of PSNR and SSIM for different GANs.}
	\label{tab:alternatives}
	\resizebox{0.7\textwidth}{!}
	{
		\begin{tabular}{c|cc|cc}
			\toprule
			& \multicolumn{2}{c|}{T1w} & \multicolumn{2}{c}{T2w} \\
			Method & PSNR & SSIM (\%) & PSNR & SSIM (\%) \\
			\midrule
			CycleGAN & 22.6$\pm$1.1 & 74.2$\pm$2.8 & 21.8$\pm$1.0 & 75.3$\pm$2.2\\
			Pix2Pix  & 23.0$\pm$1.3 & 76.2$\pm$3.4 & 22.9$\pm$0.9 & 77.2$\pm$2.8\\
			WGAN     & 24.1$\pm$1.2 & 79.4$\pm$2.4 & 23.4$\pm$0.9 & 79.5$\pm$2.0\\
			{\bf MGAN}&{\bf 26.4$\pm$0.9}&{\bf 84.0$\pm$2.2} &{\bf 25.5$\pm$0.7}&{\bf 84.8$\pm$1.8}\\
			\bottomrule
		\end{tabular}
	}
\end{table}

\subsection{Ablation Study}
Here, we investigate the effectiveness of three components of MGAN --- the SFT block, quality-guided learning, and the hybrid loss function. 
The influence of frequency transform on longitudinal prediction was verified using two variants of the metamorphic generator: (i) incorporating the SFT block equipped with both the frequency and spatial transform branches, and (ii) replacing the SFT block with a conventional spatial transform branch. We also investigated the efficacy of quality-guided learning by conducting experiments with/without quality maps generated by the discriminators. The configurations are summarized as follows:
\begin{itemize}
	\item Backbone: SFT with only spatial transform branch and without quality guidance.
	\item SFT-NCG: SFT without quality guidance. 
	\item ST-CG: Conventional spatial transform and quality guidance. 
	\item MGAN: SFT and quality guidance. 
\end{itemize}

Table~\ref{tab:ablation} indicates that MGAN achieves the highest PSNR and SSIM with a significant improvement ($p<0.05$, paired $t$-test). SFT-NCG and ST-CG perform better than Backbone, validating that wavelet-based feature mapping and quality-guidance improve prediction accuracy.

\begin{table}[!htpb]
	\renewcommand{\arraystretch}{1.3}
	\centering
	\caption{Ablation study with different MGAN configurations.}
	\label{tab:ablation}
	\resizebox{0.7\textwidth}{!}
	{
		\begin{tabular}{c|cc|cc}
			\toprule
			& \multicolumn{2}{c}{T1w} & \multicolumn{2}{c}{T2w}\\
			Model & PSNR & SSIM (\%) & PSNR & SSIM (\%) \\
			\midrule
			Backbone & 24.9$\pm$0.5 & 80.3$\pm$1.3 & 24.2$\pm$0.4 & 81.1$\pm$1.2 \\
			SFT-NCG   & 25.2$\pm$1.0 & 83.7$\pm$2.0 & 25.1$\pm$0.9 & 83.0$\pm$1.6 \\
			ST-CG & 25.9$\pm$1.2 & 82.5$\pm$2.0 & 24.8$\pm$1.0 & 82.2$\pm$1.5 \\
			\bf{MGAN} &{\bf 26.4$\pm$0.9}&{\bf 84.0$\pm$2.2} &{\bf 25.5$\pm$0.7}&{\bf 84.8$\pm$1.8} \\
			\bottomrule
		\end{tabular}
	}
\end{table} 

\begin{figure*}[!htbp]
	\begin{adjustbox}{center}
		\begin{subfigure}{1.1\textwidth}
			\includegraphics[width=\linewidth]{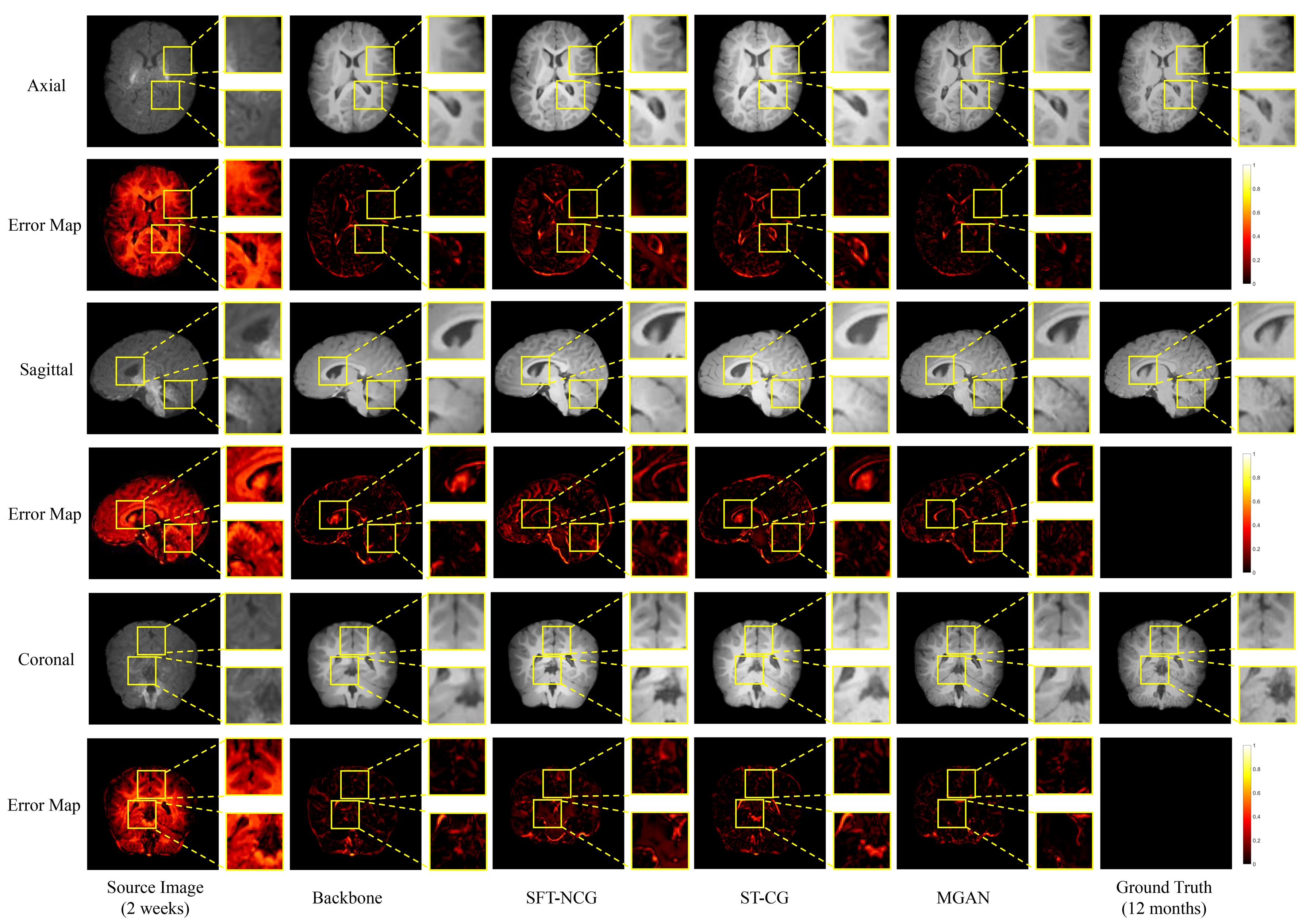}
			\caption{T1w image predictions}
			\label{fig:ablationT1}
		\end{subfigure}
	\end{adjustbox}
	
	\vspace{\baselineskip}
	
	\begin{adjustbox}{center}
		\begin{subfigure}{1.1\textwidth}
			\centering
			\includegraphics[width=\linewidth]{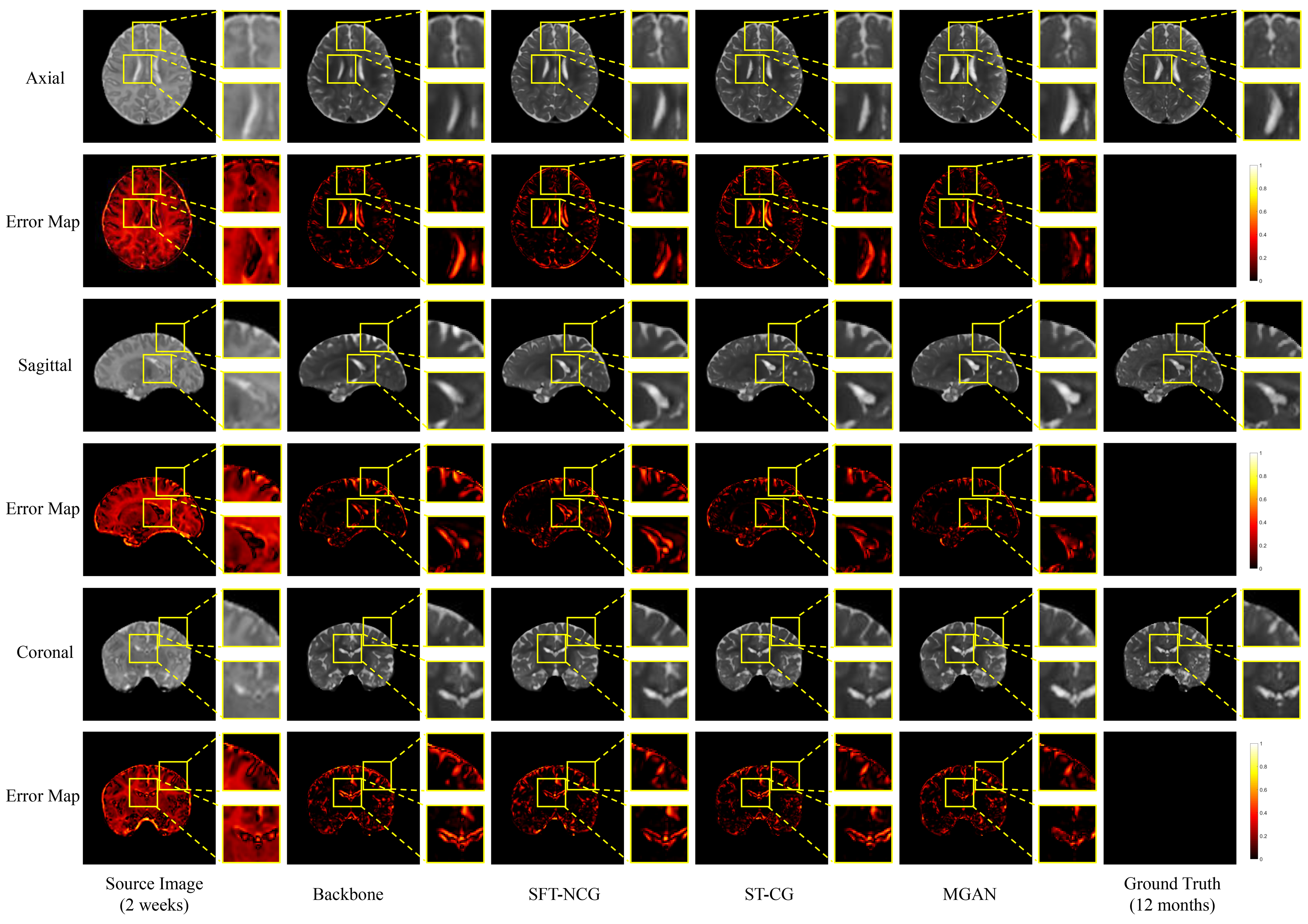}
			\caption{T2w image predictions}
			\label{fig:ablationT2}
		\end{subfigure}
	\end{adjustbox}
	\caption{Longitudinal image prediction results obtained with different MGAN configurations.}\label{fig:ablation}
\end{figure*}

Fig.~\ref{fig:ablation} shows that Backbone and ST-CG predict the 12-month scan poorly due to the spatial complexity of the cortical ribbon. SFT-NCG generated unsatisfactory results at difficult-to-predict regions as indicated by the high values in the error map. 
MGAN yields the most accurate prediction, which matches the ground truth both in terms of tissue contrast and anatomical structure.

We investigated the contribution of the uncertainty-aware loss $\mathcal{L}_{\mathrm{Q}}$, texture loss $\mathcal{L}_{\mathrm{T}}$, and frequency loss $\mathcal{L}_{\mathrm{F}}$. 
Table~\ref{tab:pairedloss} indicates that including all loss terms (Eq.~\ref{eq:gen}) yields the highest PSNR and SSIM. 
In contrast, using only the uncertainty-aware loss yields the lowest PSNR and SSIM. This implies that both the texture and frequency losses improve the predictive power of the generator.

\begin{table}[!htpb]
	\renewcommand{\arraystretch}{1.3}
	\centering
	\caption{Ablation study with different combinations of losses.}
	\label{tab:pairedloss}
	\resizebox{0.89\textwidth}{!}
	{
		\begin{tabular}{ccc|cc|cc}
			\toprule
			& & & \multicolumn{2}{c}{T1w} 
			& \multicolumn{2}{c}{T2w}\\
			$\mathcal{L}_{\mathrm{Q}}$ & $\mathcal{L}_{\mathrm{T}}$ & $\mathcal{L}_{\mathrm{F}}$ & PSNR & SSIM (\%) & PSNR & SSIM (\%)\\
			\midrule
			\checkmark&          & & 26.0$\pm$1.3 & 83.3$\pm$1.4 & 25.1$\pm$0.8 & 83.1$\pm$1.4 \\
			\checkmark&\checkmark& & 26.2$\pm$1.0 & 83.5$\pm$2.0 & 25.3$\pm$0.9 & 83.7$\pm$1.9\\
			\checkmark& &\checkmark& 26.1$\pm$1.0 & 83.7$\pm$1.8 & 25.2$\pm$0.8 & 84.3$\pm$1.7\\
			\checkmark & \checkmark & \checkmark & {\bf 26.4$\pm$0.9}&{\bf 84.0$\pm$2.2} & {\bf 25.5$\pm$0.7}&{\bf 84.8$\pm$1.8} \\
			\bottomrule
		\end{tabular}
	}
\end{table}

\begin{figure}[!t]
	\begin{adjustbox}{center}
		\begin{subfigure}{1.4\textwidth}
			\includegraphics[width=\linewidth]{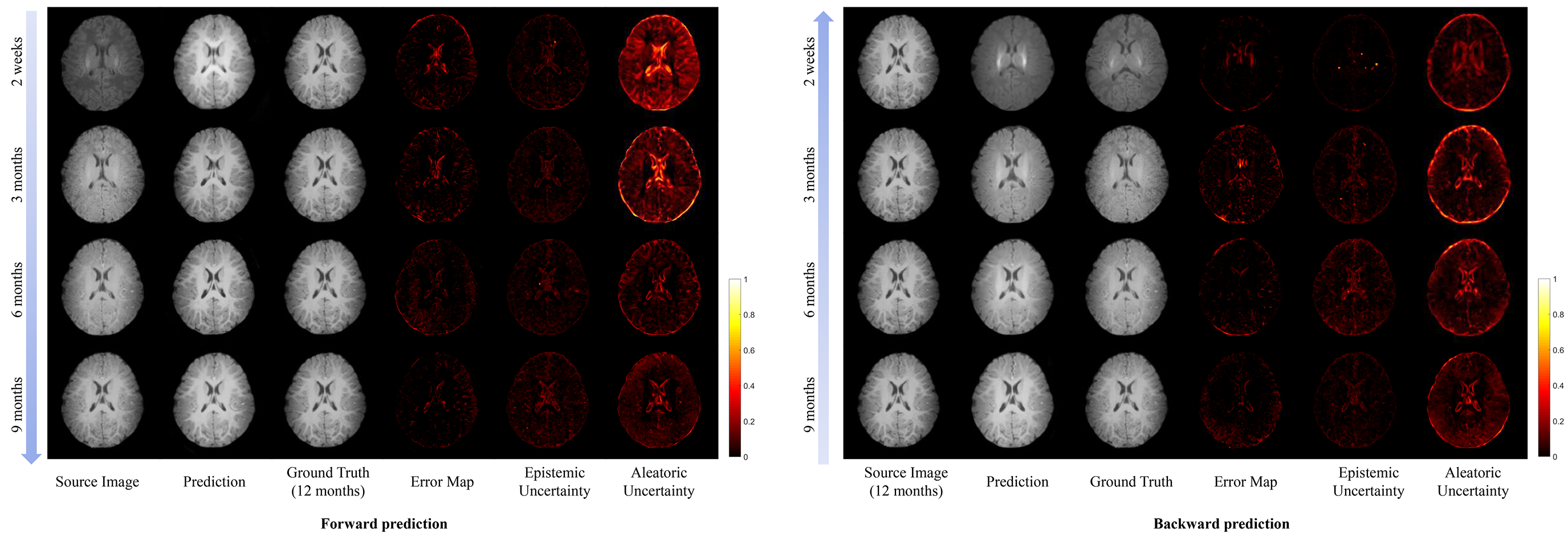}
			\caption{T1w image predictions}
			\label{fig:longitudinalT1}
		\end{subfigure}
	\end{adjustbox}
	
	\vspace{\baselineskip}
	
	\begin{adjustbox}{center}
		\begin{subfigure}{1.4\textwidth}
			\centering
			\includegraphics[width=\linewidth]{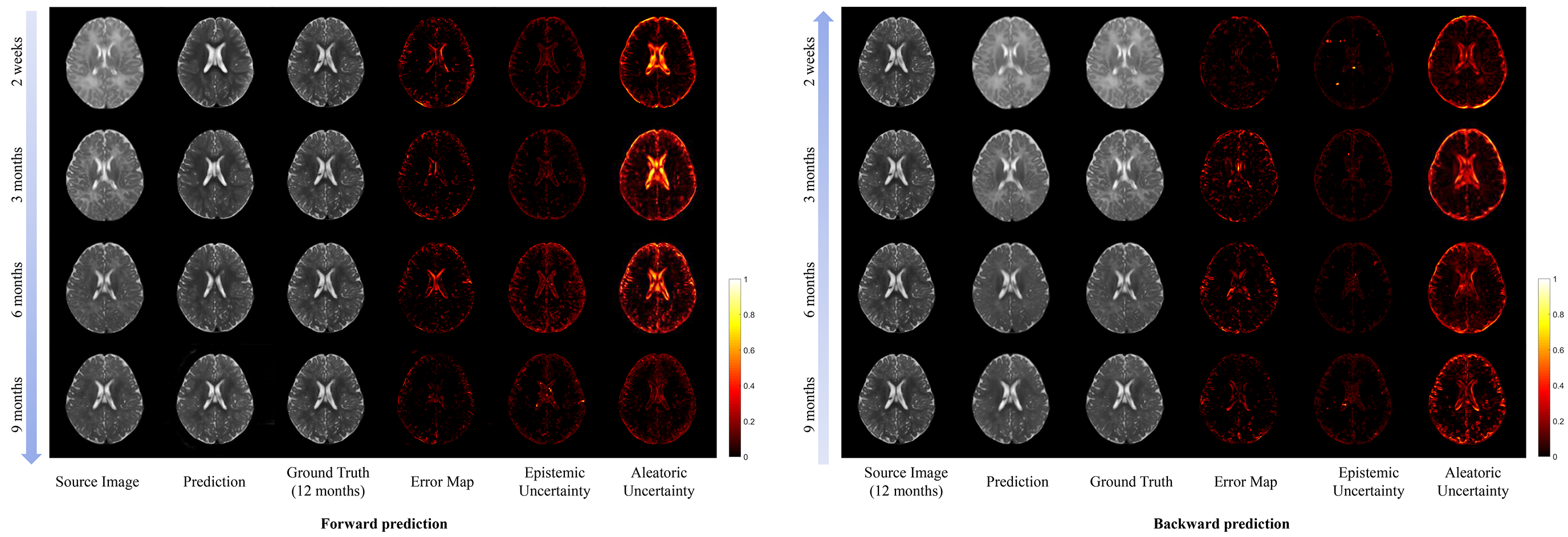}
			\caption{T2w image predictions}
			\label{fig:longitudinalT2}
		\end{subfigure}
	\end{adjustbox}
	\caption{Longitudinal prediction results for different time points. (\textit{Left}) The forward path predicts a 12-month-old image from images at earlier time points. (\textit{Right}) The backward path predicts  images of earlier time points from a 12-month-old image.}\label{fig:longitudinal}
\end{figure}

\subsection{Longitudinal Prediction}
We demonstrate the effectiveness MGAN in predicting a 12-month-old image from any earlier time-point, i.e., 2 weeks, 3 months, 6 months, and 9 months. Predictions from the forward and backward prediction paths are evaluated. 
The predicted images along with the error maps and uncertainty maps are shown in Fig.~\ref{fig:longitudinal}. The quantitative results are presented in Table~\ref{tab:longitudinal}. 
Despite the significant differences in appearance and structure, MGAN is able to predict the images with great resemblance to the ground-truth images in both tissue contrast and anatomical structure. This is validated by the high PSNR and SSIM values.
The corresponding epistemic and aleatoric uncertainty maps of the predictions are also shown in Fig.~\ref{fig:longitudinal}. 
The epistemic and aleatoric uncertainty is positively correlated with prediction errors.

\begin{table}[!htbp]
	\centering
	\renewcommand{\arraystretch}{1.3}
	\caption{Statistical summary of evaluation metrics for longitudinal prediction.}
	\label{tab:longitudinal}
	\resizebox{0.89\textwidth}{!}
	{
		\begin{tabular}{c|c|cc|cc}
			\toprule
			& &\multicolumn{2}{c}{T1w} 
			& \multicolumn{2}{c}{T2w} \\
			& & PSNR & SSIM(\%) & PSNR & SSIM(\%)\\
			\midrule
			\multirow{4}{*}{\rotatebox[origin=c]{90}{{\centering\arraybackslash{forward prediction}}}}
			& 0m$\rightarrow$12m & 26.4$\pm$0.9 & 84.0$\pm$2.2 & 25.5$\pm$0.7 & 84.8$\pm$1.8 \\
			& 3m$\rightarrow$12m & 26.1$\pm$1.3 & 84.7$\pm$4.0 & 25.7$\pm$0.9 & 83.8$\pm$2.2\\
			& 6m$\rightarrow$12m & 27.7$\pm$2.2 & 89.1$\pm$3.6 & 26.8$\pm$1.8 & 87.5$\pm$2.8 \\ 
			& 9m$\rightarrow$12m & 29.0$\pm$2.9 & 89.9$\pm$4.6 & 28.5$\pm$1.9 & 88.3$\pm$2.8 \\
			\midrule
			\multirow{4}{*}{\rotatebox[origin=c]{90}{{\centering\arraybackslash{backward prediction}}}}
			& 12m$\rightarrow$0m & 27.1$\pm$0.9 & 86.7$\pm$0.2 & 26.5$\pm$1.1 & 86.7$\pm$1.2 \\
			& 12m$\rightarrow$3m & 26.9$\pm$1.7 & 86.9$\pm$3.1 & 26.4$\pm$1.2 & 86.4$\pm$2.2\\
			& 12m$\rightarrow$6m & 27.8$\pm$1.7 & 89.7$\pm$2.7 & 27.1$\pm$1.8 & 89.4$\pm$3.2 \\ 
			& 12m$\rightarrow$9m & 28.4$\pm$2.5 & 90.5$\pm$3.1 & 27.6$\pm$2.2 & 90.1$\pm$3.4 \\
			\bottomrule
		\end{tabular}
	}
\end{table}

\begin{figure*}[!tbp]
	\centerline{\includegraphics[width=0.9\textwidth]{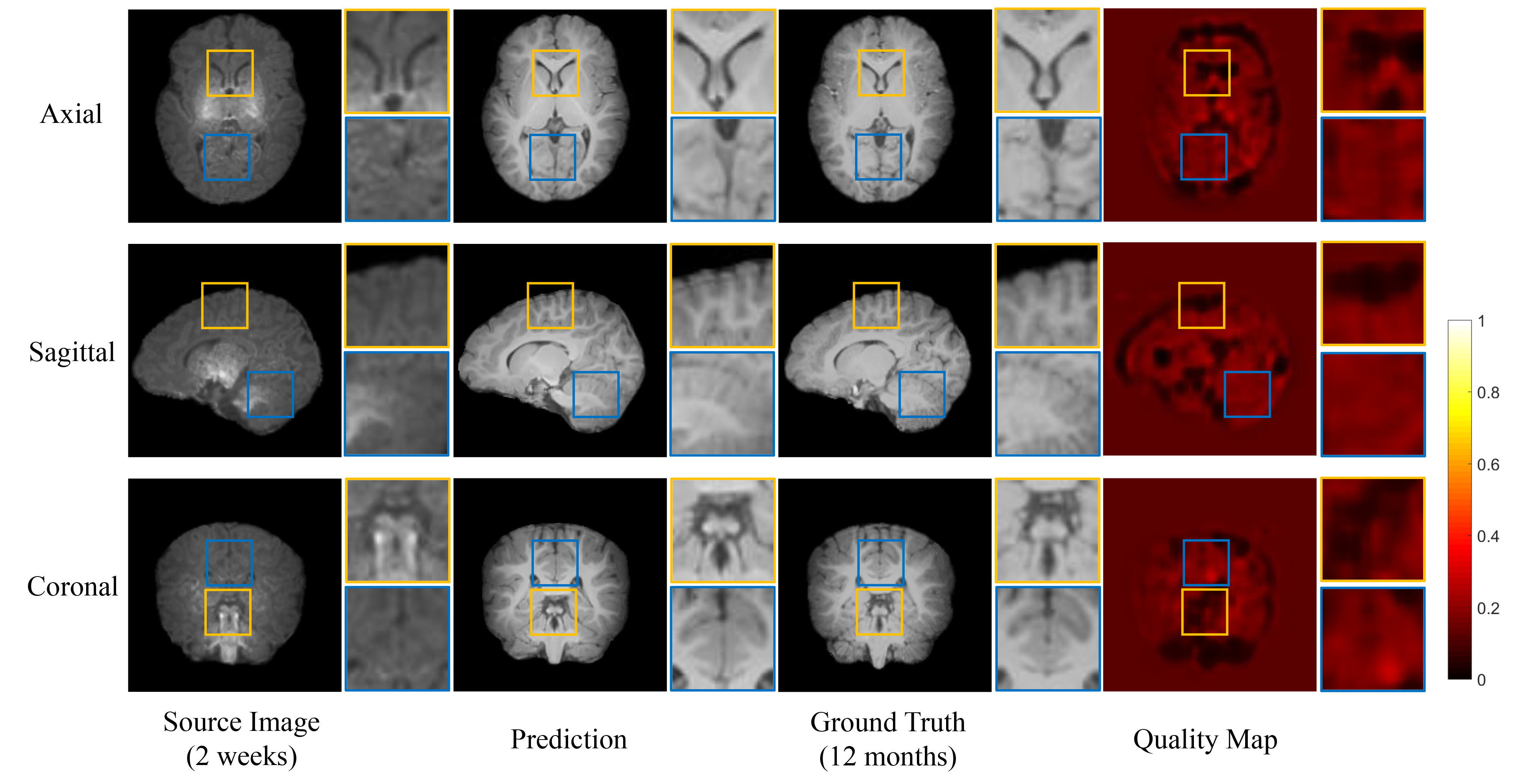}}
	\caption{Quality visualization for the 0-to-12-month-old prediction.}
	\label{fig:quality}
\end{figure*}

\section{Discussion}
\label{sec:discussion}
In this paper, we presented a metamorphic GAN that can be trained to predict infant brain MRI from one time point to another. Longitudinal prediction of infant brain MRI is challenging owing to rapid contrast and structural changes in the first year of life. 
To capture these changes, our MGAN incorporates a SFT block and integrates quality-guided learning via a hybrid loss function. 

We compared our method with existing generative adversarial networks, such as CycleGAN, Pix2Pix, and WGAN. We found that these networks are effective in prediction structures at a global scale but are less effective in predicting fine-scale structural details, especially in the cortex (Fig.~\ref{fig:alternatives}). 
In contrast, our prediction network capture spatially heterogeneous changes by employing both spatial and frequency transforms to generate feature maps. 
Particularly, DWT-based frequency transform decomposes the image into low and high frequency components to help the translation of image contrast and subtle details (Fig.~\ref{fig:ablation}). 

The quality-guided learning strategy involves using an estimation map for characterizing voxel-wise prediction quality.
Fig.~\ref{fig:quality} shows that regions with complex structures, e.g., the cerebral cortex, are associated with higher bias values. 
In contrast, regions with simple structure, e.g., lateral ventricles, are associated with lower bias values.
As shown in Fig.~\ref{fig:ablation}, employing the quality-driven $\mathcal{L}_{Q}$ loss results in more accurate predictions at challenging regions with complex patterns.
Additionally, the wavelet decomposition and gram matrix enhance the similarity between predictions and ground truths both in terms of content and style (Table~\ref{tab:pairedloss}).

\section{Conclusion}
\label{sec:conclusion}
We have proposed a trustworthy learning-based framework for longitudinal postnatal brain MRI prediction. 
The key feature our method is the utilization of wavelet transform to enable image prediction at multiple frequencies. 
We utilize quality guidance to strengthen the learning of prediction of challenging regions. 
We employ a hybrid loss function and a multi-scale discriminator to capture differences in  global intensity, style, and structure.
Experimental results demonstrate that our method achieves superior performance over several state-of-the-art image-to-image translation networks.
Despite the effectiveness of our method, it is currently trained with paired data. In future, it can be extended to be trainable with unpaired data.

\section*{Acknowledgments}
This work was supported in part by United States National Institutes of Health (NIH) grants EB008374, EB006733, and AG053867. Y. Huang was supported by the China Scholarship Council and the National Natural Science Foundation of China under Grant 6210011424.

%
%


\bibliography{refs}

\end{document}